\documentclass[preprint,nofootinbib,3p]{elsarticle}
\usepackage{graphicx}
\usepackage{amsmath}
\usepackage{amssymb}
\usepackage{overpic,subfigure}
\usepackage{multirow}
\usepackage[symbol]{footmisc}
\usepackage{dcolumn}
\usepackage{bm}
\usepackage{rotating}
\usepackage{hyperref}
\hypersetup{backref,
pdfpagemode=FullScreen,
colorlinks=true}
\usepackage{indentfirst}
\usepackage{lineno}
\usepackage{epstopdf}
\usepackage{units}
\usepackage{amsthm}
\usepackage{url}
\usepackage{amsfonts}
\usepackage{textcomp}
\usepackage{multicol}
\usepackage{verbatim}
\usepackage{rotating}
\usepackage{float}
\usepackage{color}
\usepackage{pstricks}
\usepackage{pst-node}
\usepackage{times}
\usepackage{ulem}
\usepackage{booktabs}
\usepackage{multicol}
\biboptions{numbers,sort&compress}

\newcommand{\dedx}{dE/dx}

\newcommand{\pip}{\pi^+}
\newcommand{\pim}{\pi^-}
\newcommand{\piz}{\pi^0}

\newcommand{\bfg}{\begin{figure}}
\newcommand{\efg}{\end{figure}}
\newcommand{\bitm}{\begin{itemize}}
\newcommand{\eitm}{\end{itemize}}
\newcommand{\bnum}{\begin{enumerate}}
\newcommand{\enum}{\end{enumerate}}
\newcommand{\btbl}{\begin{table}}
\newcommand{\etbl}{\end{table}}
\newcommand{\btbu}{\begin{tabular}}
\newcommand{\etbu}{\end{tabular}}

\newcommand{\beq}{\begin{equation}}
\newcommand{\edq}{\end{equation}}

\newcommand{\gev}{GeV}

\makeatletter
\newenvironment{tablehere}
  {\def\@captype{table}}
  {}
\newenvironment{figurehere}
  {\def\@captype{figure}}
  {}
\makeatother

\def\pip{\pi^{+}}
\def\pim{\pi^{-}}
\def\piz{\pi^{0}}

\def\ee{e^{+}e^{-}}
\def\dedx{\mathrm{d}E/\mathrm{d}x}

\def \gev  {\mbox{GeV}}
\def \gevc {\mbox{GeV/$c$}}
\def \gevcc{\mbox{GeV/$c^2$}}

\def \ifb  {\mbox{fb$^{-1}$}}

\def \mbc {M_{\rm{BC}}}

\def \dE {\Delta E}

\def \romanOne   {\uppercase\expandafter{\romannumeral1}}
\def \romanTwo   {\uppercase\expandafter{\romannumeral2}}
\def \romanThree {\uppercase\expandafter{\romannumeral3}}
\def \romanFour  {\uppercase\expandafter{\romannumeral4}}
\def \romanFive  {\uppercase\expandafter{\romannumeral5}}
\def \romanSix   {\uppercase\expandafter{\romannumeral6}}
\def \romanSeven {\uppercase\expandafter{\romannumeral7}}
\def \romanEight {\uppercase\expandafter{\romannumeral8}}
\def \romanNine {\uppercase\expandafter{\romannumeral9}}

\newcommand{\lamcplamcm}{\Lambda_{c}^{+}\bar{\Lambda}_{c}^{-}}
\newcommand{\lambdacp}{\Lambda_{c}^{+}}
\newcommand{\lambdacm}{\bar{\Lambda}_{c}^{-}}
\def\kshort{K^0_{\mathrm{S}}}

 \newcommand{\br}{\mathcal{B}}
 
 \newcommand{\mode}[1]{
 	\ifnum#1=1
 	\lambdacp \rightarrow \Lambda \pip \pim e^{+} \nu_e
 	\else
 	\ifnum#1=2
 	\lambdacp \rightarrow p \kshort \pim e^{+} \nu_e
 	\fi
 	\fi
 }
 
  \newcommand{\state}[1]{
 	\ifnum#1=1
 	\Lambda \pip \pim e^{+} \nu_e
 	\else
 	\ifnum#1=2
 	p \kshort \pim e^{+} \nu_e
 	\fi
 	\fi
 }

\journal{Physics Letters B}

\begin{document}
\begin{frontmatter}
\title{\boldmath Search for the semi-leptonic decays $\Lambda_c^+ \to \Lambda \pi^+ \pi^- e^+ \nu_e$ and $\Lambda_c^+ \to p K_S^0 \pi^- e^+ \nu_e$}

\author{
\begin{small}
\begin{center}
M.~Ablikim$^{1}$, M.~N.~Achasov$^{13,b}$, P.~Adlarson$^{73}$, R.~Aliberti$^{34}$, A.~Amoroso$^{72A,72C}$, M.~R.~An$^{38}$, Q.~An$^{69,56}$, Y.~Bai$^{55}$, O.~Bakina$^{35}$, I.~Balossino$^{29A}$
% ...
\\
\vspace{0.2cm}
(BESIII Collaboration)\\
\vspace{0.2cm} {\it
$^{1}$ Institute of High Energy Physics, Beijing 100049, People's Republic of China\\
% ...
$^{13}$ G.I. Budker Institute of Nuclear Physics SB RAS (BINP), Novosibirsk 630090, Russia\\
% ...
$^{29}$ INFN Sezione di Ferrara, (A)INFN Sezione di Ferrara, I-44122, Ferrara, Italy; (B)University of Ferrara,  I-44122, Ferrara, Italy\\
% ...
$^{34}$ Johannes Gutenberg University of Mainz, Johann-Joachim-Becher-Weg 45, D-55099 Mainz, Germany\\
$^{35}$ Joint Institute for Nuclear Research, 141980 Dubna, Moscow region, Russia\\
% ...
$^{38}$ Liaoning Normal University, Dalian 116029, People's Republic of China\\
% ...
$^{55}$ Southeast University, Nanjing 211100, People's Republic of China\\
$^{56}$ State Key Laboratory of Particle Detection and Electronics, Beijing 100049, Hefei 230026, People's Republic of China\\
% ...
$^{69}$ University of Science and Technology of China, Hefei 230026, People's Republic of China\\
% ...
$^{72}$ University of Turin and INFN, (A)University of Turin, I-10125, Turin, Italy; (B)University of Eastern Piedmont, I-15121, Alessandria, Italy; (C)INFN, I-10125, Turin, Italy\\
$^{73}$ Uppsala University, Box 516, SE-75120 Uppsala, Sweden\\
% ...
$^{b}$ Also at the Novosibirsk State University, Novosibirsk, 630090, Russia\\
% ...
}\end{center}

\vspace{0.4cm}
\end{small}
}

\date{\today}
\begin{abstract}

We search for the semi-leptonic decays $\Lambda_c^+ \to \Lambda \pi^+ \pi^- e^+ \nu_e$ and $\Lambda_c^+ \to p K_S^0 \pi^- e^+ \nu_e$ in a sample of  4.5\,$\mathrm{fb}^{-1}$ of $e^{+}e^{-}$ annihilation data collected in the center-of-mass energy region between 4.600\,GeV and 4.699\,GeV by the BESIII detector at the BEPCII. No significant signals are observed, and the upper limits on the decay branching fractions are set to be $\mathcal{B}(\Lambda_c^+ \to \Lambda \pi^+ \pi^- e^+ \nu_e)<3.9\times10^{-4}$ and $\mathcal{B}(\Lambda_c^+ \to p K_S^0 \pi^- e^+ \nu_e)<3.3\times10^{-4}$ at the 90\% confidence level, respectively.

\end{abstract}

\end{frontmatter}
\begin{multicols}{2}

\setrunninglinenumbers
	

\section{Introduction}


The study of $\lambdacp$ semi-leptonic (SL) decays provides valuable information about weak and strong interactions in baryons containing a heavy quark. According to Fermi's Golden Rule, the decay rate depends on the product of kinematic phase space (PHSP) and dynamic amplitude. The dynamic amplitude in $\lambdacp$ SL decays is much simpler than in non-leptonic decays, and can be factorized into a hadronic term, leptonic term and weak quark-mixing Cabibbo-Kobayashi-Maskawa matrix element~\cite{CKM}. The hadronic current describes the weak transition of the charm quark to a light quark, and the leptonic current describes the coupling to the charged-lepton--neutrino pair. In principle, the leptonic current can be precisely calculated, in contrast to the hadronic current, which suffers from difficulties due to the strong interaction~\cite{Asner:2008nq}, and can be parameterized by form factors. Recently, the first measurement of $\lambdacp\to\Lambda$ form factors from the BESIII Collaboration~\cite{BESIII:2022ysa} shows large discrepancies with the lattice Quantum Chromodynamics (LQCD) calculation~\cite{Meinel:2016dqj}, which attracts wide attention.


In Ref.~\cite{BESIII:2022ysa}, improved measurement of the absolute branching fraction (BF) $\br(\lambdacp\to\Lambda e^+\nu_e)=(3.56\pm0.11_{\mathrm{stat.}}\pm0.07_{\mathrm{syst.}})\%$ was also reported. Comparing this result with the BF of $\lambdacp$ inclusive SL decay $\br(\lambdacp\to X e^+\nu_e)=(3.95\pm0.34_{\mathrm{stat.}}\pm0.09_{\mathrm{syst.}})$~\cite{BESIII:2018mug} indicates that there is still potential room for other exclusive SL decays of the order of $10^{-4}$ to $10^{-3}$. Ref.~\cite{Gronau:2018vei} suggests these other decay modes could be from $\lambdacp$ decays to excited states such as $\Lambda(1405)$ and $\Lambda(1520)$ or continuum $\Sigma\pi$ and $NK$ contributions. Recently, BESIII presented the first observation of the SL decay $\lambdacp\to pK^- e^+ \nu_e$~\cite{BESIII:2022qaf}, in which evidence for $\lambdacp\to\Lambda(1520)e^+\nu_e$ is reported with a BF of $(1.02\pm0.52_{\mathrm{stat.}}\pm0.11_{\mathrm{syst.}})\times10^{-3}$ and a combined statistical and systematic significance of 3.3$\sigma$. This result stimulates further research on $\Lambda^+_c$ SL decays into various excited $\Lambda^*$ baryons. Over the years, many theoretical calculations concerning $\lambdacp\to\Lambda^{*}$ form factors and BFs based on constituent quark model (CQM)~\cite{Pervin:2005ve}, nonrelativistic quark model (NRQM)~\cite{Hussain:2017lir}, light-front quark model (LFQM)~\cite{Li:2021qod} and LQCD~\cite{Meinel:2021grq} have been performed; their results for the BF of $\lambdacp\to\Lambda^{*}e^+\nu_e$ are shown in Table~\ref{tab:bf theory}. Searching for $\Lambda^+_c$ SL decays which may contribute to these excited $\Lambda^*$ baryons is important for testing and constraining these theoretical calculations.

In this paper, we search for the SL decays $\mode{1}$ and $\mode{2}$, with the final state of $p\pip\pim\pim e^{+}\nu_{e}$, based on 4.5 $\ifb$ of $\ee$ annihilation data collected at the center-of-mass system (CMS) energies $\sqrt{s}$ = 4.600, 4.612, 4.628, 4.641, 4.661, 4.682, and 4.699\,GeV~\cite{BESIII:2015qfd,BESIII:2015zbz,BESIII:2022xii,BESIII:2022ulv} by the BESIII detector at the BEPCII collider. Throughout this paper, charge-conjugate modes are implied.

\begin{tablehere}
	\centering
	\caption{The BFs for $\lambdacp\to\Lambda^{*}e^{+}\nu_e$ predicted by different theoretical models, in units of $10^{-4}$.\label{tab:bf theory}}
	\begin{scriptsize}
	\begin{tabular}{ccccc}
		\toprule
		$\Lambda^{*}$ state & CQM~\cite{Pervin:2005ve} & NRQM~\cite{Hussain:2017lir} & LFQM~\cite{Li:2021qod} & LQCD~\cite{Meinel:2021grq} \\
		\midrule
		$\Lambda(1520)$ & $10.00$ & $5.94$                       & -----                                  & $5.12\pm0.82$ \\
		$\Lambda(1600)$ & $4.00$   & $1.26$                       & $(0.7\pm0.2)$   & ----- \\
		$\Lambda(1890)$ & -----        & $3.16\times10^{-2}$  & -----                                  & ----- \\
		$\Lambda(1820)$ & -----        & $1.32\times10^{-2}$  & -----                                  & ----- \\
		\bottomrule
	\end{tabular}
	\end{scriptsize}
\end{tablehere}


\section{BESIII experiment and Monte Carlo simulation}

The BESIII detector~\cite{Ablikim:2009aa} records symmetric $e^+e^-$ collisions provided by the BEPCII storage ring~\cite{Yu:2016cof} in the CMS energy range from 2.0 to 4.95\,GeV, with a peak luminosity of $1 \times10^{33}\;\text{cm}^{-2}\text{s}^{-1}$ achieved at $\sqrt{s} = 3.77\;\text{GeV}$. The cylindrical core of the BESIII detector covers 93\% of the full solid angle and consists of a helium-based multilayer drift chamber~(MDC), a plastic scintillator time-of-flight system (TOF), and a CsI(Tl) electromagnetic calorimeter (EMC), which are all enclosed in a superconducting solenoidal magnet providing a 1.0\,T magnetic field~\cite{Huang:unity}. The solenoid is supported by an octagonal flux-return yoke with resistive plate counter muon identification modules interleaved with steel. The charged-particle momentum resolution at $1\,\gevc$ is $0.5\%$, and the specific ionization energy loss $\dedx$ resolution is $6\%$ for electrons from Bhabha scattering. The EMC measures photon energies with a resolution of $2.5\%$ ($5\%$) at $1\,\gev$ in the barrel (end-cap) region. The time resolution in the TOF barrel region is 68\,ps, while that in the end-cap region is 110\,ps. The end-cap TOF system was upgraded in 2015 using multi-gap resistive plate chamber technology, providing a time resolution of 60\,ps~\cite{li:TOF}.

Simulated data samples are produced with a {\sc Geant4}-based~\cite{GEANT4:2002zbu} Monte Carlo (MC) package, which includes the geometric description of the BESIII detector~\cite{GDMLMethod,BesGDML} and the detector response. The simulation models the beam-energy spread and initial-state radiation (ISR) in the $\ee$ annihilations with the generator {\sc kkmc}~\cite{kkmc}. The final-state radiation (FSR) from charged final-state particles is incorporated using {\sc photos}~\cite{photos}.

The inclusive MC sample includes the production of $\lamcplamcm$ pairs, and open-charm mesons, ISR production of vector charmonium(-like) states, and continuum processes which are incorporated in {\sc kkmc}~\cite{kkmc, kkmc2}. Known decay modes are modeled with {\sc evtgen}~\cite{evtgen, besevtgen} using the BFs taken from the Particle Data Group (PDG)~\cite{ParticleDataGroup:2022pth}. The signal decay modes $\mode{1}$ and $\mode{2}$ are not included in the inclusive MC sample. The remaining unknown charmonium decays are modeled with {\sc lundcharm}~\cite{lundcharm, Yang:2014vra}. The inclusive MC sample is used to study background contributions and to optimize event selection criteria.

The $\ee\to\lamcplamcm$ signal MC sample are generated to estimate the detection efficiencies, in which the $\lambdacp$ decays through signal modes while the other $\lambdacm$ decays through 12 single-tag (ST) modes as described below. In the baseline analysis, all $\Lambda\pip\pim$ combinations are assumed to be from the $\Lambda(1520)$ resonance. The SL decay of $\lambdacp \to \Lambda(1520) e^+ \nu_e$ is simulated based on the heavy-quark effective-theory (HQET) model~\cite{Hussain:2017lir}, while the decay of $\Lambda(1520) \to \Lambda\pip\pim$ is simulated with the uniformly distributed PHSP model. The SL decay $\mode{2}$ is simulated by using the PHSP model. For two-body $\lambdacm$ ST modes, the angular distributions are described with the transverse polarization and decay asymmetry parameters of the $\Lambda_{c}^{+}$ and its daughter baryons~\cite{Ronggang:Lambda_c}. For three-body and four-body $\lambdacm$ ST modes, the intermediate states are modeled according to individual internal partial-wave analysis models. 


\section{Event selection}

At the CMS energy region between $\sqrt{s}$ = 4.600\,GeV and 4.699\,GeV, the $\lambdacp\lambdacm$ pair is produced in the electron-positron annihilation without additional hadron companions. Since the neutrinos in the signal decays can not be detected by the BESIII detector, we use the double tag (DT) technique which was first applied by the Mark III Collaboration~\cite{MARK-III:1985hbd}.

The ST sample consists of events in which the $\lambdacm$ baryon is reconstructed with any of the following 12 exclusive hadronic decay modes: $\lambdacm\to \bar{p}\kshort$, $\bar{p}K^{+}\pi^-$, $\bar{p}\kshort\piz$, $\bar{p}\kshort\pim\pip$, $\bar{p}K^+\pim\piz$, $\bar{\Lambda}\pim$, $\bar{\Lambda}\pim\piz$, $\bar{\Lambda}\pim\pip\pim$, $\bar{\Sigma}^{0}\pim$, $\bar{\Sigma}^{-}\piz$, $\bar{\Sigma}^{-}\pim\pip$ and $\bar{p}\pim\pip$, which are the same as those considered in Ref.~\cite{BESIII:2022xne}. The DT sample is formed of those events in the ST sample that also contain candidates for the $\mode{1}$ and $\mode{2}$ decays. The BF of the signal mode $s$ ($s=\Lambda\pi\pi$ corresponds to $\mode{1}$ and $s=p\kshort\pi$ corresponds to $\mode{2}$) is determined by
\begin{equation}
	\br_{s} = \frac{N_{s}^{\mathrm{DT}}}{ \sum_{i}N^{\mathrm{ST}}_{i}\cdot \varepsilon_{s,i}^{\mathrm{DT}}/\varepsilon_{i}^{\mathrm{ST}} \cdot \br_{s}^{\mathrm{inter}} } = 
	\frac{N_{s}^{\mathrm{DT}}}{ N^{\mathrm{ST}} \cdot \varepsilon_{s}^{\mathrm{sig}}\cdot \br_{s}^{\mathrm{inter}}},
\label{equ:BF_DT}
\end{equation}
where $N_{s}^{\mathrm{DT}}$ is the DT signal yield of the signal-mode $s$, $N^{\mathrm{ST}}_{i}$ is the ST event yield of tag-mode $i$, $\varepsilon_{i}^{\mathrm{ST}}$ is the efficiency of detecting a tag-mode $i$ candidate, $\varepsilon_{s,i}^{\mathrm{DT}}$ is the efficiency of simultaneously detecting the tag-mode $i$ and the signal-mode $s$ candidate, $\br_{\Lambda\pi\pi}^{\mathrm{inter}}=\br(\Lambda\to p\pim)$ and $\br_{p\kshort\pi}^{\mathrm{inter}}=\br(\kshort\to\pip\pim)$ are the BFs of the decays of the intermediate states taken from the PDG\cite{ParticleDataGroup:2022pth}, $N^{\mathrm{ST}}=\sum_{i}N^{\mathrm{ST}}_{i}$ is the total yield of all 12 ST modes in data, and $\varepsilon_{s}^{\mathrm{sig}} = \frac{\sum_{i}N^{\mathrm{ST}}_{i}\cdot \varepsilon_{s,i}^{\mathrm{DT}}/\varepsilon_{i}^{\mathrm{ST}}}{\sum_{i}N^{\mathrm{ST}}_{i}}$ is the average efficiency of detecting a $\lambdacp$ decaying into the signal-mode $s$ in the system recoiling against the ST $\lambdacm$.

The selection criteria of the ST $\lambdacm$ candidate events follow the previous BESIII analysis~\cite{BESIII:2022xne}. The beam-energy-constrained mass $\mbc\equiv\sqrt{E^2_{\mathrm{beam}}/c^4-|\vec{p}|^2/c^2}$ is used to identify the ST $\lambdacm$ candidates, where $E_{\mathrm{beam}}$ is the average value of the $e^+$ and $e^-$ beam energies, $\vec{p}$ is the total measured $\lambdacm$ momentum in the CMS of the $\ee$ collision. To improve the signal purity, the energy difference $\dE\equiv E-E_{\mathrm{beam}}$ for the $\lambdacm$ candidate is required to fulfill a mode-dependent $\dE$ requirement. Here, $E$ is the total reconstructed energy of the $\lambdacm$ candidate in the CMS of the $\ee$ collision. For each ST mode, if more than one candidate satisfies the above requirements, the one with the minimal $|\dE|$ is retained. The total yield of the 12 ST modes is $N^{\mathrm{ST}}=123509\pm461$, where the uncertainty is statistical. Full information about the $\dE$ requirements, $\mbc$ distributions, signal regions, ST yields, and efficiencies for the various ST modes at each energy point is detailed in Ref.~\cite{BESIII:2022xne}.

The signal candidate events for $\mode{1}$ and $\mode{2}$ are selected with those tracks in the event that are not used to form the ST $\lambdacm$ candidates. The $\Lambda$ and $\kshort$ candidates are reconstructed from $p\pim$ and $\pip\pim$ combinations, respectively. Charged tracks are reconstructed in the MDC, and are required to have a polar angle $\theta$ with respect to the $z$-axis, defined as the symmetry axis of the MDC, satisfying $|\!\cos\theta|<0.93$. The distance of closest approach to the interaction point (IP) is required to be less than 10~cm along the $z$-axis ($V_z$) and less than 1~cm in the perpendicular plane ($V_r$), which are denoted as tight track requirements. Tracks originating from $\kshort$ and $\Lambda$ decays are not subjected to these distance requirements. Instead, they are subjected to the loose track requirements of $|V_z|<20\,\mathrm{cm}$ and no restriction on $V_r$. To suppress background events, it is required that there are only five charged tracks to be reconstructed in the signal side, which must satisfy loose track requirements. 

Particle identification (PID)~\cite{Ping:2009zzc} for charged tracks is implemented using combined information from the flight time measured in the TOF and the $\dedx$ measured in the MDC. Charged tracks are identified as protons when they satisfy $\mathcal{L}(p)> \mathcal{L}(K)$, $\mathcal{L}(p)> \mathcal{L}(\pi)$ and $\mathcal{L}(p)>0$, where $\mathcal{L}(h)$ is the PID probability for each particle $(h)$ hypothesis with $h=p,\pi,K$. Charged tracks are identified as pions when they satisfy $\mathcal{L}(\pi) > \mathcal{L}(K)$ and $\mathcal{L}(\pi)>0$. The energy deposited in the EMC is also considered when constructing the PID probability for the positron hypothesis, $\mathcal{L}(e)$. Charged tracks are identified as positrons when they satisfy $\mathcal{L}(e)> 0.001$ and a requirement on the PID probability ratio which is $\mathcal{L}(e)/(\mathcal{L}(e)+\mathcal{L}(\pi)+\mathcal{L}(K))>0.99(0.98)$ for $\mode{1}$ ($\mode{2}$). The energy loss due to FSR and bremsstrahlung photon(s) of the positron candidates is partially recovered by adding the showers that are within a $5^{\circ}$ cone relative to the track momentum.

Long-lived $\Lambda$ ($\kshort$) candidates are reconstructed by combining $p\pim$ ($\pip\pim$) pairs. Here, a PID requirement is imposed on the proton candidate, but not on the $\pi$ candidates. A vertex fit is applied to pairs of charged tracks, constraining them to originate from a common decay vertex, and the $\chi^2$ of this vertex fit is required to be less than 100. The invariant mass of the $p\pim$ ($\pip\pim$) pair must satisfy $1.09<M_{p\pim}<1.14\,\gevcc$ ($0.490<M_{\pip\pim}<0.504\,\gevcc$). Here, $M_{p\pim}$ and $M_{\pip\pim}$ are the invariant masses of $p\pim$ and $\pip\pim$ pairs, calculated with the common decay-vertex constraint imposed. To further suppress background, we require a positive value of the decay length. These selection criteria are optimized by using the Punzi figure-of-merit (FOM)~\cite{Punzi:2003bu}. The definition of the Punzi FOM is $\varepsilon/(3/2+\sqrt{B})$, where $\varepsilon$ is the detection efficiency and $B$ is the number of background events in the inclusive MC sample. If there are more than one $\Lambda~(\kshort$) or $e^{+}$ candidates in an event, the $\Lambda$($\kshort$) with the largest $L/\sigma_{L}$ or the $e^+$ with the largest $\mathcal{L}(e)$ is retained to avoid double counting of the DT events.

The missing energy and missing momentum carried by undetected neutrinos are denoted by $E_{\mathrm{miss}}$ and $\vec{p}_{\mathrm{miss}}$, which are calculated from $E_{\mathrm{miss}} = E_{\mathrm{beam}} - E_{\mathrm{SL}}$ and $\vec{p}_{\mathrm{miss}} = \vec{p}_{\lambdacp} - \vec{p}_{\mathrm{SL}}$ in the initial $\ee$ rest frame. Here, $E_{\mathrm{SL}}$ and $\vec{p}_{\mathrm{SL}}$ are the measured energy and momentum of SL decay products, which are determined as $E_{\mathrm{SL}} = E_{\Lambda(p)} + E_{\pip(\kshort)} + E_{\pim} + E_{e}$ and $\vec{p}_{\mathrm{SL}} = \vec{p}_{\Lambda(p)} + \vec{p}_{\pip(\kshort)} + \vec{p}_{\pim} + \vec{p}_{e}$. The momentum $\vec{p}_{\lambdacp}$ is given by $\vec{p}_{\lambdacp} = -\hat{p}_{\mathrm{tag}}\sqrt{E^{2}_{\mathrm{beam}}/c^{2}-m^{2}_{\lambdacm}c^{2}}$, where $\hat{p}_{\mathrm{tag}}$ is the direction of the momentum of the ST $\lambdacm$ and $m_{\lambdacm}$ is the known mass of the $\lambdacm$~\cite{ParticleDataGroup:2022pth}. If the ST $\lambdacm$ and SL decay products in the signal side are correctly identified, the $U_{\mathrm{miss}} = E_{\mathrm{miss}} - c|{\vec{p}_{\mathrm{miss}}}|$ is expected to peak around zero for the signal mode. Signal candidates are required to satisfy $U_{\mathrm{miss}}\in [-0.08,0.08]\,\gev$, which is about three times the resolution evaluated in the simulation.

With the help of a generic event type analysis tool, TopoAna~\cite{Zhou:2020ksj}, the inclusive MC sample is used to study background events after applying the primary selection criteria described above.
In the events of the ST modes $\lambdacm \to \bar{p} \pip\pim$ and $\lambdacm \to \bar{\Sigma}^-\pip\pim$, to suppress contamination due to misidentification of pions and positrons, only positrons with a detected energy deposit in the EMC are retained.
The background levels of these two modes are the highest among the 12 tag modes due to the random combination of final state particles, and this additional selection allows to enhance the discrimination of the PID. To suppress $\gamma$-conversion background events, $\cos\theta_{e,\pi}$ is required to be less than $0.88(0.92)$ for $\mode{1}$ ($\mode{2}$), where $\theta_{e,\pi}$ is the opening angle between the oppositely charged pion and positron. To suppress contamination from $\lambdacp\to\Lambda\pip\pim\pip\,(p\kshort\pip\pim)$ decays, the invariant mass $M_{\Lambda\pip\pim e(\pi)^+}\, (M_{p\kshort\pim e(\pi)^+})$, where $e(\pi)^+$ denotes that the $e^+$ mass is replaced by that of the $\pi^+$ in the calculation, is required to be less than $2.20\, (2.28)\,\gevcc$. The remaining dominant background events are from $\lambdacp \to \Lambda\pip\omega/\eta$ with $\omega/\eta\to\pip\pim\piz$, and $\lambdacp \to \Sigma^0\pip\pim\pip$ with $\Sigma^0\to\gamma\Lambda$ for $\mode{1}$, and $\lambdacp \to p\kshort\eta$ with $\eta\to\pip\pim\piz$ or $\eta\to\gamma\ee$ for $\mode{2}$. In these events, the charged pions misidentified with positrons and the electromagnetic showers due to $\pi^0/\gamma$ are not detected. To suppress this category of background, $\cos\theta_{P_{\mathrm{miss}},\gamma}$ is required to be less than $0.82(0.90)$ for $\mode{1}$ ($\mode{2}$), where $\theta_{P_{\mathrm{miss}},\gamma}$ is the opening angle between the missing momentum $P_{\mathrm{miss}}$ and the most energetic shower. After applying these requirements, which have been optimized through the Punzi FOM, the level of background is greatly suppressed. The resulting $U_{\mathrm{miss}}$ distributions of candidates in data and MC samples are shown in Figure~\ref{fig:umiss}, where no significant excess over the expected backgrounds is observed. The total number of observed events, $N^{\mathrm{obs}}$, is counted in the $U_{\mathrm{miss}}$ signal region and listed in Table~\ref{tab:variable_likelihood}. The average efficiencies $\varepsilon_{\Lambda\pi\pi}^{\mathrm{sig}}$ and $\varepsilon_{p\kshort\pi}^{\mathrm{sig}}$ are determined to be (9.69\;$\pm$\;0.03)\% and (13.58\;$\pm$\;0.02)\%, respectively, where the uncertainties are statistical. The two dimensional efficiency maps of $M(\Lambda\pip\pim)$ or $M(p\kshort\pim)$ versus $q^2$ are shown in Figure~\ref{fig:2D_eff}, where $q^2=(p_{e}+p_{\nu_e})^2$.

\begin{table*}[htbp]
	\centering
	\caption{The total number of observed events $N^{\mathrm{obs}}$ in the signal region, the average efficiency $\varepsilon^{\mathrm{sig}}$, the number of events in the SB region $N^{\mathrm{SB}}_{\mathrm{bkg1}}$, the number of $bkg2$ events estimated by MC simulation $N^{\mathrm{MC}}_{\mathrm{bkg2}}$, the corresponding statistical uncertainty $\sigma^{\mathrm{MC}}_{\mathrm{bkg2}}$ for $\mode{1}$ and $\mode{2}$ and the upper limit on the DT signal yield $N^{\mathrm{DT}}$ at the 90\% confidence level. The uncertainties are statistical.\label{tab:variable_likelihood}}
	\small
	\begin{tabular}{ cccccc }
		\toprule
		Decay mode & $N^{\mathrm{obs}}$ & $\varepsilon^{\mathrm{sig}}$ (\%) & $N^{\mathrm{SB}}_{\mathrm{bkg1}}$ & $N^{\mathrm{MC}}_{\mathrm{bkg2}}\pm\sigma^{\mathrm{MC}}_{\mathrm{bkg2}}$ & $N^{\mathrm{DT}}$ \\
		\midrule
		$\mode{1}$ & 3 & 9.69\;$\pm$\;0.03 & 9 & 4.8\;$\pm$\;0.4 & 2.9 \\
		$\mode{2}$ & 2 & 13.58\;$\pm$\;0.02 & 0 & 2.2\;$\pm$\;0.3 & 3.8 \\
		\bottomrule
	\end{tabular}
\end{table*}

The backgrounds can be separated into two categories: events with a wrong ST candidate denoted as $bkg1$ which is dominantly from non-$\Lambda_c$ decay process, and events with a correct ST but wrong signal candidate denoted as $bkg2$ which is dominantly from $\Lambda_c$ decay process. The size of the $bkg1$ component can be estimated with the surviving events in the ST sideband (SB) region of $M_{\mathrm{BC}}^{\mathrm{tag}}$, which is defined as $(2.25,2.27)\,\gevcc$. The corresponding number of $bkg1$ events, $N_{\mathrm{bkg1}}$, is estimated from the number of events in the SB region $(N^{\mathrm{SB}}_{\mathrm{bkg1}})$ normalized by a scale factor $r$, which is the ratio of the integrated numbers of background events in the signal region and SB region. The scale factor $r$ is found to be $1.533\pm0.004$, where the uncertainty is statistical only. The number of events in the SB region, $N^{\mathrm{
SB}}_{\mathrm{bkg1}}$, is expected to follow a Poisson ($\mathcal{P}$) distribution with central value of $N_{\mathrm{bkg1}}\times\frac{1}{r}$. The size of the $bkg2$ component can be estimated with the inclusive $\lambdacp\lambdacm$ MC sample by subtracting the wrong ST events, and the corresponding number of events, $N_{\mathrm{bkg2}}$, is expected to follow a Gaussian function ($\mathcal{G}$), with central value $N^{\mathrm{MC}}_{\mathrm{bkg2}}$ and standard deviation $\sigma^{\mathrm{MC}}_{\mathrm{bkg2}}$. The relevant numbers are summarized in Table~\ref{tab:variable_likelihood}.

\begin{figurehere}
	\centering
	\subfigure[]
	{\includegraphics[width=6.5cm]{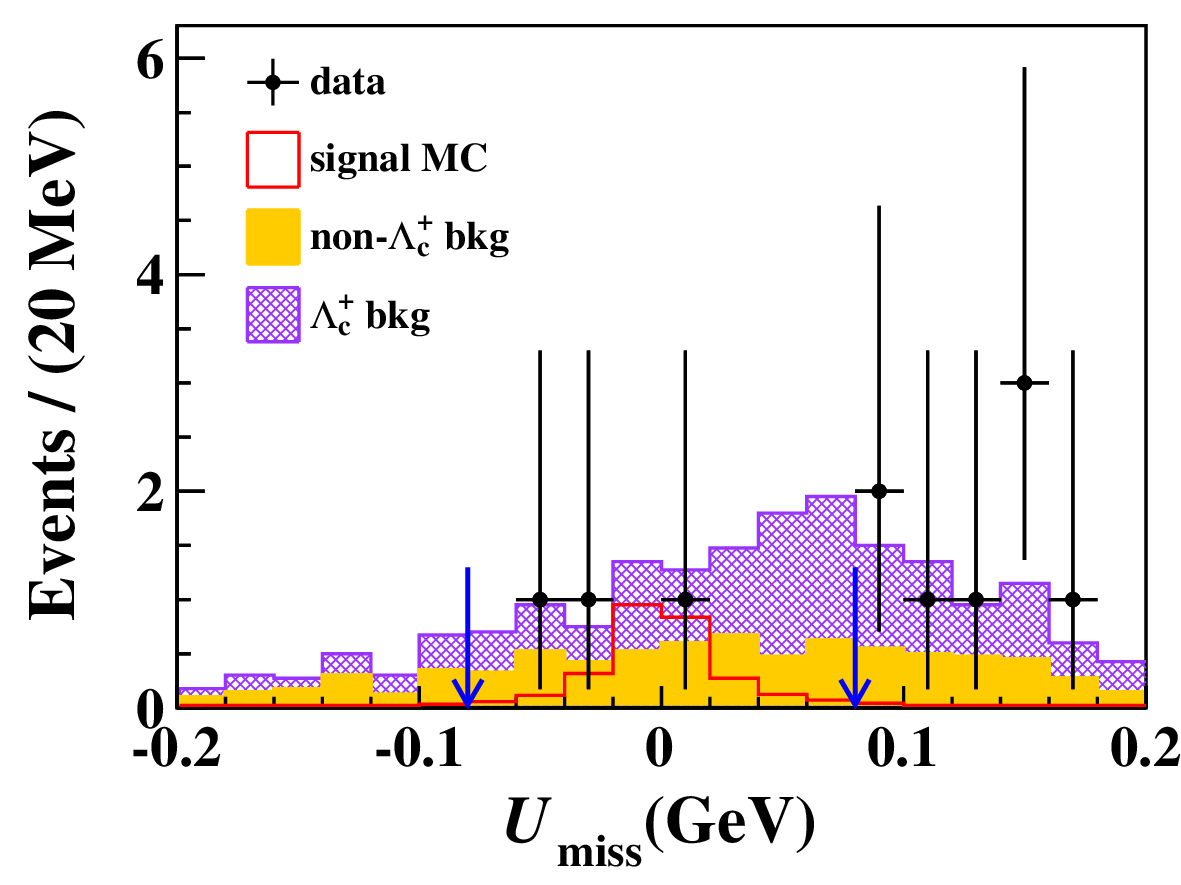}}
	\subfigure[]
	{\includegraphics[width=6.5cm]{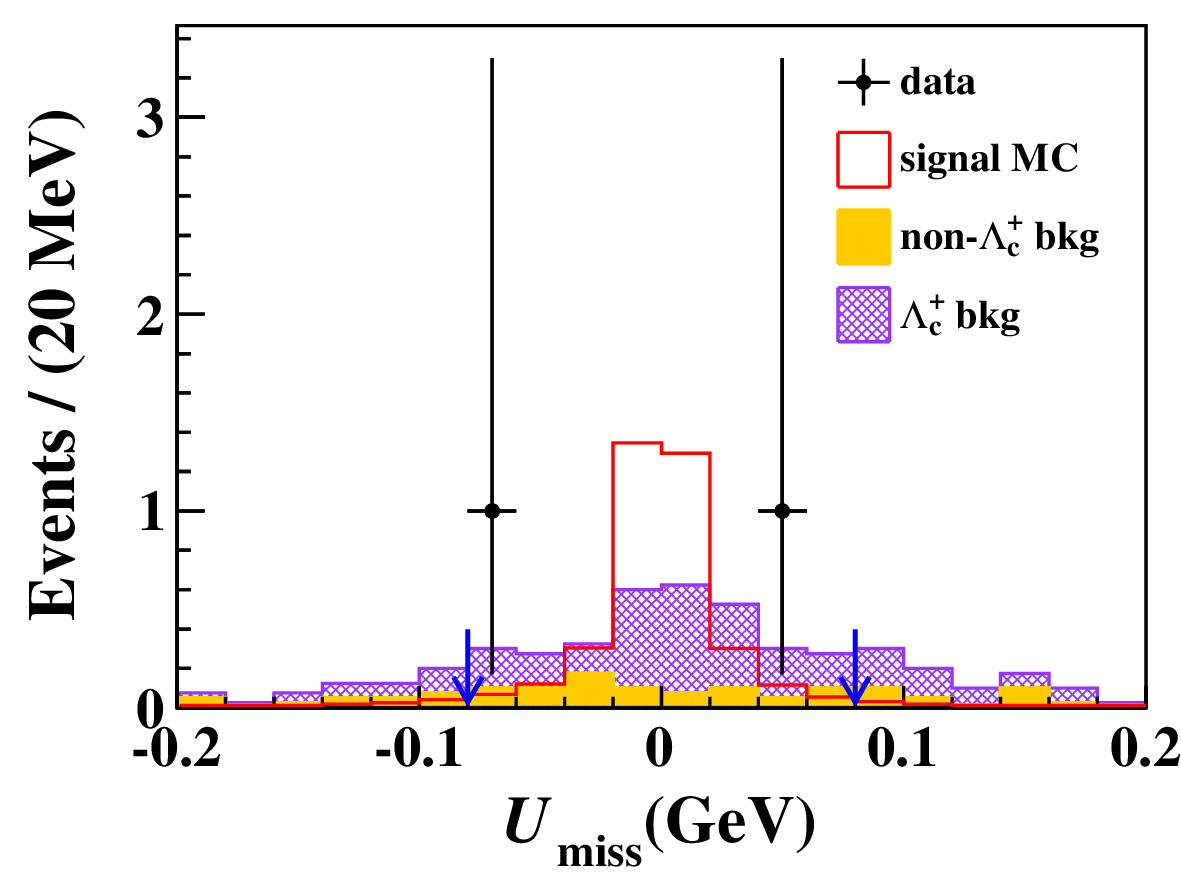}}
	\caption{The $U_{\mathrm{miss}}$ distributions of candidates for (a) $\state{1}$ and (b) $\state{2}$, where the black points with error bars denote data, the violet hatched histogram the $\lambdacp$ decay background, the orange solid histogram the non-$\lambdacp$ decay background and the red hollow histogram the signal MC sample, which is scaled to the measured upper limit on the signal yield at 90\% confidence level. The region between two arrows indicates the signal region.\label{fig:umiss}}
\end{figurehere}

\begin{figurehere}
	\centering
	\subfigure[]
	{\includegraphics[width=6.5cm]{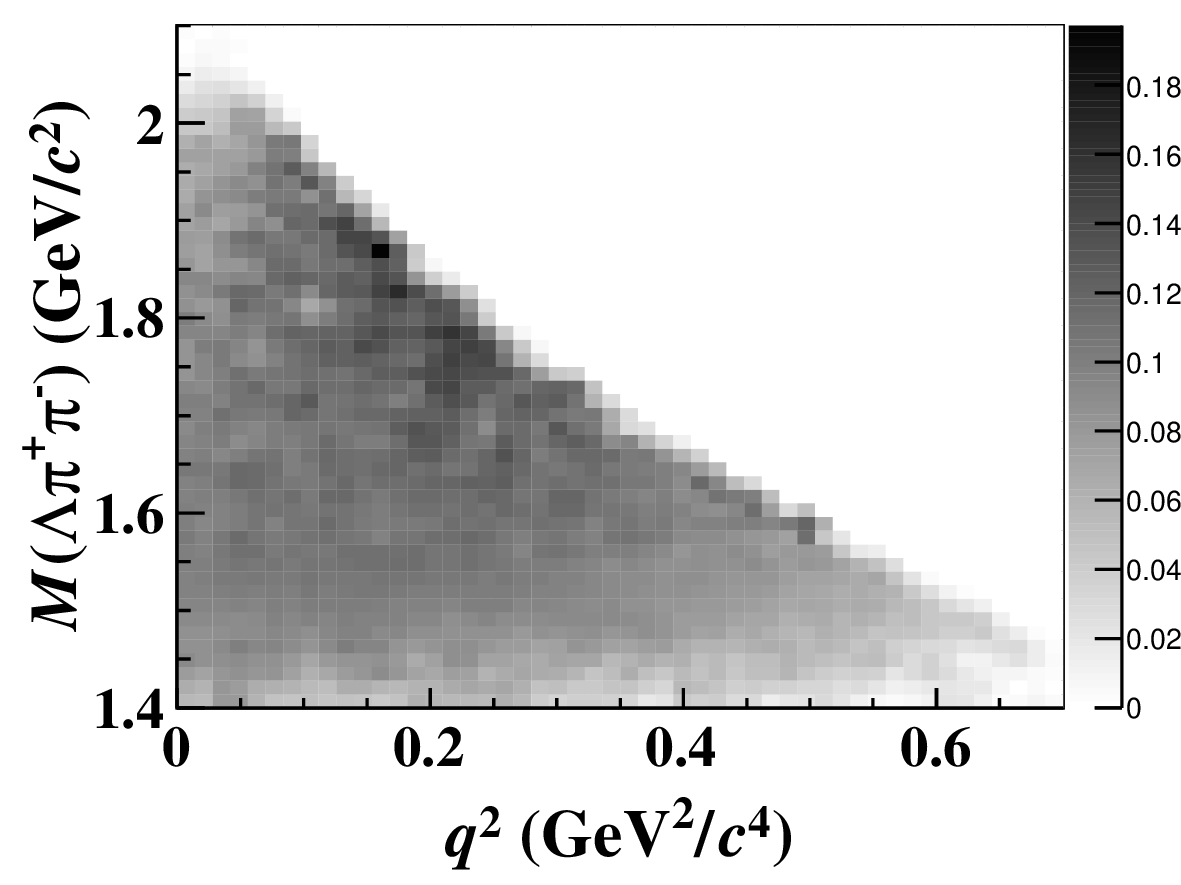}}
	\subfigure[]
	{\includegraphics[width=6.5cm]{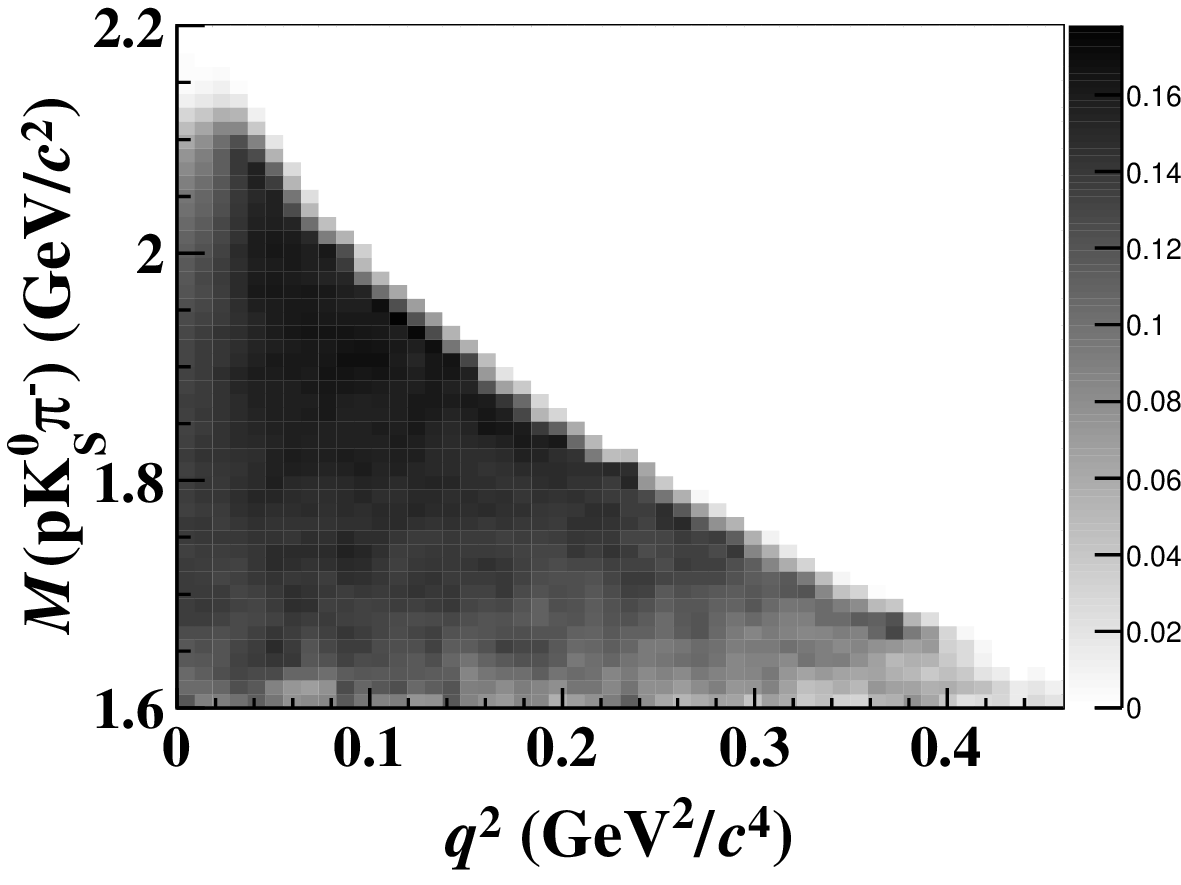}}
	\caption{The two dimensional efficiency maps for (a) $\state{1}$ and (b) $\state{2}$ of $M(\Lambda\pip\pim)$ or $M(p\kshort\pim)$ versus $q^2$. In order to show the model independent efficiency map, for $\state{1}$, PHSP signal MC samples mixed with different $\Lambda^{*}$ states are used here.\label{fig:2D_eff}}
\end{figurehere}


\section{Systematic uncertainties}

With the DT technique, the systematic uncertainties in the BF measurements due to the detection and reconstruction of the ST $\lambdacm$ baryons mostly cancel out. For the signal side, the signal yield, $N_{\mathrm{sig}}$, which is $N_{\Lambda\pi\pi}^{\mathrm{DT}}$ or $N_{p\kshort\pi}^{\mathrm{DT}}$, is calculated by $N^{\mathrm{obs}} - N_{\mathrm{bkg1}} - N_{\mathrm{bkg2}}$, where $N^{\mathrm{obs}}$ is the total number of observed events obtained from counting, without any uncertainty assigned, $N_{\mathrm{bkg1}}$ and $N_{\mathrm{bkg2}}$ are the numbers of $bkg1$ and $bkg2$ events, the statistical uncertainties of which are assigned assuming Poisson and Gaussian distributions, respectively. All sources of systematic uncertainties are summarized in Table~\ref{tab:syst summary} and discussed below. It should be noted that the systematic uncertainties due to the $M_{\Lambda\pip\pim e(\pi)^+}$ or $M_{p\kshort\pim e(\pi)^+}$ requirement are negligible.

\begin{center}
\begin{tablehere}
	\centering
	\caption{Relative systematic uncertainties for the measurements of the BFs of $\mode{1}$ and $\mode{2}$. The total systematic uncertainty is the sum in quadrature of the individual components. ``-----'' indicates the cases where there is no uncertainty.\label{tab:syst summary}}
	\small
	\begin{tabular}{ lcc }
		\toprule
		Source & $\br_{\Lambda\pi\pi}$(\%) &  $\br_{p\kshort\pi}$(\%) \\
		\midrule
		MC sample size & 0.3 & 0.2\\
		Number of ST $\lambdacm$ & 0.4 & 0.4\\
		$\br^{\mathrm{inter}}$ & 0.8 & 0.1\\
		$p$ tracking & ----- & 0.4\\
		$p$ PID & ----- & 0.2\\
		$\pi$ tracking & 2.6 & 0.4\\
		$\pi$ PID & 0.7 & 0.3\\
		$e$ tracking & 0.5 & 0.1\\
		$e$ PID & 2.8 & 3.6\\
		$\Lambda$ reconstruction & 2.2 & -----\\
		$\kshort$ reconstruction & ----- & 3.2\\
		$\cos\theta_{e,\pi}$~requirement & 1.5 & 1.5\\
		$\cos\theta_{P_{\mathrm{miss}},\gamma}$~requirement & 0.1 & 0.1\\
		FSR recovery & 0.2 & 0.2\\
		Signal model & 2.2 & 5.6\\
		\midrule
		Total & 5.2 & 7.5\\
		\bottomrule
	\end{tabular}
\end{tablehere}
\end{center}

(\romanOne) \emph{MC sample size}. The statistical uncertainties arising from the MC are propagated as systematic uncertainties, which are 0.3\% and 0.2\% for $\br_{\Lambda\pi\pi}$ and $\br_{p\kshort\pi}$, respectively.

(\romanTwo) \emph{Number of ST $\lambdacm$}. The statistical uncertainty on the number of ST $\lambdacm$, 0.4\%, is considered as a systematic uncertainty.

(\romanThree) \emph{$\br^{\mathrm{inter}}$}. The uncertainties on the known BFs of $\br_{\Lambda\pi\pi}^{\mathrm{inter}}$ and $\br_{p\kshort\pi}^{\mathrm{inter}}$ are 0.8\% and 0.1\% for $\br_{\Lambda\pi\pi}$ and $\br_{p\kshort\pi}$, respectively.

(\romanFour) \emph{$p$ tracking/PID}. 
The proton (anti-proton) tracking/PID efficiency is studied with a $J/\psi \to p\bar{p}\pip\pim$ control sample. The detection efficiency of $\mode{2}$ is recalculated after re-weighting the signal MC sample on an event-by-event basis according to the momentum- and polar angle-dependent efficiency differences between data and MC simulation. The relative differences between the baseline and corrected efficiencies, 0.4\% and 0.2\%, are taken as the systematic uncertainties due to $p$ tracking and PID efficiencies for $\br_{p\kshort\pi}$, respectively. The systematic uncertainties due to $p$ tracking and PID efficiencies for $\br_{\Lambda\pi\pi}$ are included in the $\Lambda$ reconstruction, as described below.

(\romanFive) \emph{$\pi$ tracking/PID}. 
The charged pion tracking/PID efficiency is also studied with the $J/\psi \to p\bar{p}\pip\pim$ control sample. The detection efficiencies of $\mode{1}$ and $\mode{2}$ are re-weighted in the same way as in the $p$ tracking study. The resultant data-MC differences are assigned as the systematic uncertainties, which are 2.6\% and 0.4\% in $\pi$ tracking, and 0.7\% and 0.3\% in $\pi$ PID for $\br_{\Lambda\pi\pi}$ and $\br_{p\kshort\pi}$, respectively.

(\romanSix) \emph{$e$ tracking}. 
The positron tracking efficiency is studied with a $\ee\to\gamma\ee$ control sample. The detection efficiencies of $\mode{1}$ and $\mode{2}$ are re-weighted in the same way as in the $p$ tracking study. The resultant data-MC differences, 0.5\% and 0.1\%, are assigned as the systematic uncertainties for $\br_{\Lambda\pi\pi}$ and $\br_{p\kshort\pi}$, respectively.

(\romanSeven) \emph{$e$ PID}. The positron PID efficiency with the requirement of $\mathcal{L}(e)>0.001$ and $\mathcal{L}(e)/(\mathcal{L}(e)+\mathcal{L}(\pi)+\mathcal{L}(K))>0.99(0.98)$ is studied with control samples of $\ee\to\gamma\ee$ and $D^{0}\to \bar{K}^{0}\pim e^{+}\nu_e$ decays. The differences in the acceptance efficiencies between data and MC simulation are assigned as the corresponding systematic uncertainties, which are 2.8\% and 3.6\% for $\br_{\Lambda\pi\pi}$ and $\br_{p\kshort\pi}$, respectively.


(\romanEight) \emph{$\Lambda$ reconstruction}. The $\Lambda$ reconstruction efficiency is studied by using the control samples of $J/\psi\to\Lambda p K^{-}$ and $J/\psi\to\Lambda\bar{\Lambda}$ decays. Using the same procedure as in the $p$ tracking study, the systematic uncertainty due to $\Lambda$ reconstruction is assigned to be 2.2\% for $\br_{\Lambda\pi\pi}$.

(\romanNine) \emph{$\kshort$ reconstruction}. The $\kshort$ reconstruction efficiency of the selections $0.490<M_{\pip\pim}<0.504\,\gevcc$ and $L/\sigma_{L}\,\textgreater\,0$ is studied by using control samples of $J/\psi\to K^{*}(892)^{\mp}K^{\pm}, K^{*}(892)^{\mp}\to\kshort\pi^{\mp}$, $J/\psi\to\phi\kshort K^{\pm}\pi^{\mp}$ and $D^{0}\to \bar{K}^{0}\pim e^{+}\nu_e$ decays. The difference in the acceptance efficiencies between data and MC simulation is assigned as the corresponding systematic uncertainty, which is 3.2\% for $\br_{p\kshort\pi}$.


(\uppercase\expandafter{\romannumeral10}) \emph{$\cos\theta_{e,\pi}$~requirement/$\cos\theta_{P_{\mathrm{miss}},\gamma}$~requirement/FSR recovery}. The systematic uncertainties due to the $\cos\theta_{e,\pi}$~requirement, $\cos\theta_{P_{\mathrm{miss}},\gamma}$~requirement and FSR recovery are assigned to be 1.5\%, 0.1\%, and 0.2\% for the two signal modes, respectively, from measuring the differences in the acceptance efficiencies between data and MC simulation with the control sample of $D^{0}\to \bar{K}^{0}\pim e^{+}\nu_e$ decays.

(\uppercase\expandafter{\romannumeral11}) \emph{Signal model}. To evaluate the systematic uncertainty due to signal model, additional $\Lambda^{*}$ resonance contributions are considered. For the two signal modes, MC events of $\Lambda^+_c\to\Lambda(1820)/\Lambda(1890)e^+\nu_e$ are generated in both the PHSP model and the HQET model~\cite{Hussain:2017lir}. The contribution from the transition through the $\Lambda(1600)$ is also considered for $\mode{1}$. The largest changes in the detection efficiencies are assigned as the associated systematic uncertainties, which are 2.2\% and 5.6\% for $\br_{\Lambda\pi\pi}$ and $\br_{p\kshort\pi}$, respectively.

\section{BF upper limits}

To calculate the upper limits on the BFs of the signal decays, we use a maximum likelihood estimator extended from the profile likelihood method~\cite{Rolke:2004mj}. According to Eq.~\eqref{equ:BF_DT}, the effective signal yield is defined to be $N^{\mathrm{eff}}$ which follows a Gaussian distribution with mean $\br^{\mathrm{inter}}\cdot N^{\mathrm{ST}}\cdot\varepsilon^{\mathrm{sig}}$, and width $\br^{\mathrm{inter}}\cdot N^{\mathrm{ST}}\cdot\varepsilon^{\mathrm{sig}}\cdot\sigma$, where $\sigma$ is the relative uncertainty of $N^{\mathrm{eff}}$ including both statistical and systematic components. From error propagation it follows that $\sigma$ is equal to the relative systematic uncertainty of $\br$, as given in Table~\ref{tab:syst summary}. Therefore, the joint likelihood is
\begin{equation}
	\begin{split}
		\mathcal{L} = 
		&\mathcal{P}(N^{\mathrm{obs}}| N^{\mathrm{eff}}\cdot\br+N_{\mathrm{bkg1}}+N_{\mathrm{bkg2}}) \\
		&\cdot \mathcal{G}(N^{\mathrm{eff}}| \br^{\mathrm{inter}}\cdot N^{\mathrm{ST}}\cdot\varepsilon^{\mathrm{sig}}, \br^{\mathrm{inter}}\cdot N^{\mathrm{ST}}\cdot\varepsilon^{\mathrm{sig}}\cdot\sigma)
		\\
		&\cdot \mathcal{P}(N^{\mathrm{SB}}_{\mathrm{bkg1}}| N_{\mathrm{bkg1}}/r)
		\\
		&\cdot \mathcal{G}(N_{\mathrm{bkg2}}| N_{\mathrm{bkg2}}^{\mathrm{MC}}, \sigma_{\mathrm{bkg2}}^{\mathrm{MC}}).
		\label{equ:likelihood}
	\end{split}
\end{equation}
Based on the Bayesian statistics, $\br$ is priorly assumed to be the uniform distribution, and the likelihood $\mathcal{L}$ maximized by the variation of the other parameters $N^{\mathrm{eff}}$, $N_{\mathrm{bkg1}}$ and $N_{\mathrm{bkg2}}$, is the posterior probability of $\br$. By scanning $\br$, the likelihood distribution as a function of $\br$ is obtained.

The resultant distributions of likelihoods plotted as a function of the individual BFs of $\mode{1}$ and $\mode{2}$ are shown in Figure~\ref{fig:UL}. The upper limits on the signal BFs at the 90\% confidence level (CL) are estimated by integrating the likelihood curves in the physical region of $\br\ge0$~\cite{Liu:2015uha}. The upper limits on the BFs of $\mode{1}$ and $\mode{2}$ are determined to be $3.9\times10^{-4}$ and $3.3\times10^{-4}$, respectively, and the related upper limit on the DT signal yield is listed in Table~\ref{tab:variable_likelihood}.

\begin{figurehere}
  \centering
  \includegraphics[width=6.5cm]{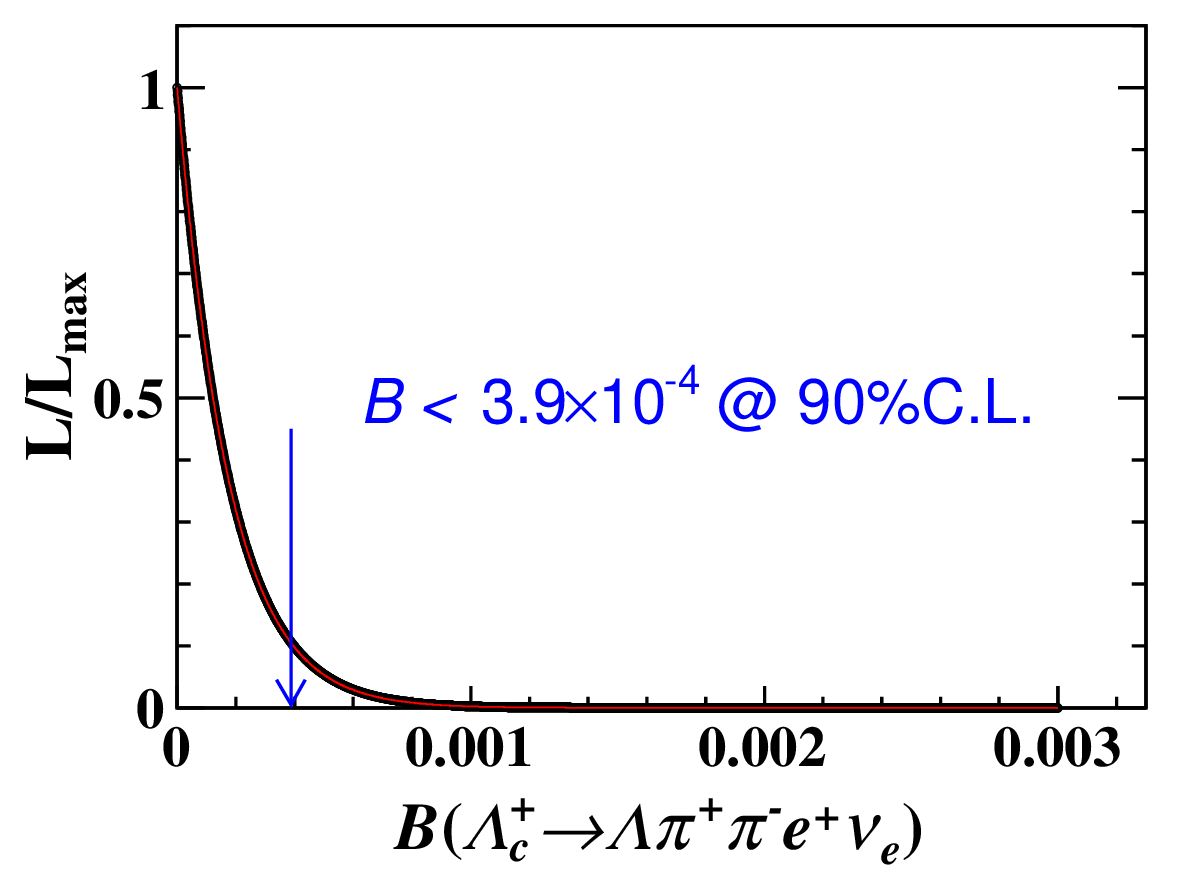}
  \includegraphics[width=6.5cm]{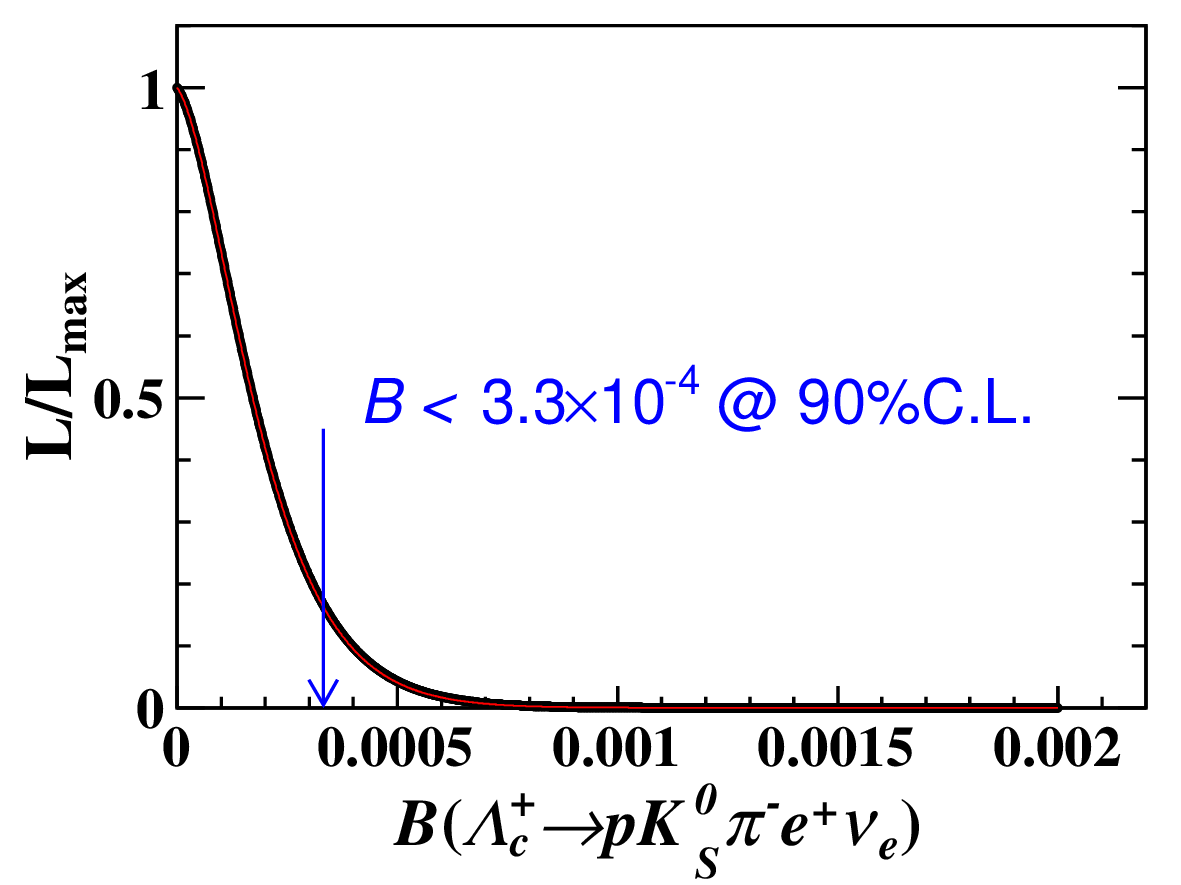}
  \caption{Normalized likelihood distribution as a function of the signal BFs. The black curve denotes the nominal fit result including systematic uncertainty, the red curve denotes the fit result without incorporating the systematic uncertainties and the arrows point to the positions of the upper limits at the 90\% CL. The two curves are totally overlapping.
  \label{fig:UL}}
\end{figurehere}

Assuming that all the $\Lambda\pi\pi$ combinations come from $\lambdacp \to \Lambda(1520)e^+\nu_e$, which is expected to be the dominant decay, the upper limit of $\mathcal{B}(\lambdacp\to\Lambda(1520)e^+\nu_e)$ is determined to be $4.3\times10^{-3}$ at 90\% CL after considering $\br(\Lambda(1520)\to\Lambda\pip\pim)=(10\pm1)\%$~\cite{ParticleDataGroup:2022pth}. Assuming that all the $\Lambda\pi\pi$ combinations come from $\lambdacp \to \Lambda(1600)e^{+}\nu_{e}$, the upper limit of $\br(\Lambda_{c}^{+}\to\Lambda(1600)e^{+}\nu_{e})$ is determined to be $9.0\times10^{-3}$ at 90\% CL after taking into account $\br(\Lambda(1600)\to\Sigma(1385)\pi)=(9\pm4)\%$ and $\br(\Sigma(1385)\to\Lambda\pi)=(87.5\pm1.5)\%$~\cite{ParticleDataGroup:2022pth}. Our result is consistent with $\br(\Lambda_{c}^{+}\to\Lambda(1520)e^{+}\nu_{e})=(1.02\pm0.52_{\mathrm{stat.}}\pm0.11_{\mathrm{syst.}})\times10^{-3}$, as measured via $\Lambda(1520)\to p K^-$ by BESIII~\cite{BESIII:2022qaf}.

\section{Summary}

In summary, based on 4.5 $\mathrm{fb}^{-1}$ of $e^{+}e^{-}$ annihilation data collected in the CMS energy region between 4.600\,GeV and 4.699\,GeV by the BESIII detector at the BEPCII, we search for the SL decays $\Lambda_c^+ \to \Lambda \pi^+ \pi^- e^+ \nu_e$ and $\Lambda_c^+ \to p K_S^0 \pi^- e^+ \nu_e$. No significant signal is observed in data. Therefore, the upper limits on the BFs of these two decays are set to be $\mathcal{B}(\Lambda_c^+ \to \Lambda \pi^+ \pi^- e^+ \nu_e)<3.9\times10^{-4}$ and $\mathcal{B}(\Lambda_c^+ \to p K_S^0 \pi^- e^+ \nu_e)<3.3\times10^{-4}$ at 90\% CL. Assuming that all the $\Lambda\pi\pi$ combinations come from $\Lambda(1520)$ or $\Lambda(1600)$, the BF upper limits are determined to be $\br(\Lambda_{c}^{+}\to\Lambda(1520)e^{+}\nu_{e}) < 4.3\times10^{-3}$ and $\br(\Lambda_{c}^{+}\to\Lambda(1600)e^{+}\nu_{e}) < 9.0\times10^{-3}$ at 90\% CL. Due to the limitation of statistics, our results are consistent with all theoretical calculations listed in Table~\ref{tab:bf theory}. The result on $\br(\Lambda_{c}^{+}\to\Lambda(1520)e^{+}\nu_{e})$ is also consistent with that measured via $\Lambda(1520)\to p K^-$ in Ref.~\cite{BESIII:2022qaf}. This result helps to constrain the theoretical calculations of the BFs and form factors of $\Lambda\to\Lambda^{*}e^+\nu_e$. The larger data samples that will be collected at BESIII in future~\cite{BESIII:2020nme,Li:2021iwf} will allow the sensitivity to these decays to be further improved, and provide a deeper understanding of charmed baryon decays.

\section*{Acknowledgments}

The BESIII Collaboration thanks the staff of BEPCII and the IHEP computing center for their strong support. This work is supported in part by National Key R\&D Program of China under Contracts Nos. 2020YFA0406400, 2020YFA0406300; National Natural Science Foundation of China (NSFC) under Contracts Nos. 11635010, 11675275, 11735014, 11835012, 11935015, 11935016, 11935018, 11961141012, 12022510, 12025502, 12035009, 12035013, 12061131003, 12175321, 12192260, 12192261, 12192262, 12192263, 12192264, 12192265, 12221005; the Chinese Academy of Sciences (CAS) Large-Scale Scientific Facility Program; the CAS Center for Excellence in Particle Physics (CCEPP); Joint Large-Scale Scientific Facility Funds of the NSFC and CAS under Contract No. U1832207, U1932101; State Key Laboratory of Nuclear Physics and Technology, PKU under Grant No. NPT2020KFY04; CAS Key Research Program of Frontier Sciences under Contracts Nos. QYZDJ-SSW-SLH003, QYZDJ-SSW-SLH040; 100 Talents Program of CAS; The Institute of Nuclear and Particle Physics (INPAC) and Shanghai Key Laboratory for Particle Physics and Cosmology; ERC under Contract No. 758462; European Union's Horizon 2020 research and innovation programme under Marie Sklodowska-Curie grant agreement under Contract No. 894790; German Research Foundation DFG under Contracts Nos. 443159800, 455635585, Collaborative Research Center CRC 1044, FOR5327, GRK 2149; Istituto Nazionale di Fisica Nucleare, Italy; Ministry of Development of Turkey under Contract No. DPT2006K-120470; National Research Foundation of Korea under Contract No. NRF-2022R1A2C1092335; National Science and Technology fund; National Science Research and Innovation Fund (NSRF) via the Program Management Unit for Human Resources \& Institutional Development, Research and Innovation under Contract No. B16F640076; Polish National Science Centre under Contract No. 2019/35/O/ST2/02907; Suranaree University of Technology (SUT), Thailand Science Research and Innovation (TSRI), and National Science Research and Innovation Fund (NSRF) under Contract No. 160355; The Royal Society, UK under Contract No. DH160214; The Swedish Research Council; U. S. Department of Energy under Contract No. DE-FG02-05ER41374.




\end{multicols}
\end{document}